\documentclass[hyper]{JHEP} 

\usepackage{epsfig}

\newcommand\fverb{\setbox\pippobox=\hbox\bgroup\verb}

\newcommand\fverbdo{\egroup\medskip\noindent%

            \fbox{\unhbox\pippobox}\ }

\newcommand\fverbit{\egroup\item[\fbox{\unhbox\pippobox}]}

\newbox\pippobox

\title{Note About T-duality of Non-Relativistic String}
\author{J. Kluso\v{n}\\
Department of
Theoretical Physics and Astrophysics\\
Faculty of Science, Masaryk University\\
Kotl\'{a}\v{r}sk\'{a} 2, 611 37, Brno\\
Czech Republic\\
E-mail: \email{klu@physics.muni.cz}} \preprint{}

 \abstract{In this note we perform canonical analysis of T-duality
    for non-relativistic string in stringy Newton-Cartan background. We confirm
    recent result that T-duality along longitudinal spatial direction of stringy Newton-Cartan geometry maps  non-relativistic
    string to the  relativistic string  that propagates on the background with
     light-like isometry.}

\def\mM{\mathcal{M}}
\def\tH{\tilde{H}}

\def\bA{\mathbf{A}}

\def\tz{\tilde{z}}

\def\blambda{\bar{\lambda}}
\def\ty{\tilde{y}}

\def\bB{\mathbf{B}}

\def\tx{\tilde{x}}

\def\be{\begin{equation}}

\def\ee{\end{equation}}

\def\bea{\begin{eqnarray}}

\def\htau{\hat{\tau}}
\def\eea{\end{eqnarray}}

\def\tY{\mathcal{Y}}

\def\mH{\mathcal{H}}

\newcommand{\tB}{\tilde{B}}

\newcommand{\mG}{\mathcal{G}}

\def \bA{\mathbf{A}}

\def\tg{\tilde{g}}

\newcommand{\mL}{\mathcal{L}}

\def \tY{\tilde{Y}}
\def\pb #1{\left\{#1\right\}}

\begin{document}
\section{Introduction and Summary}
Recently, very interesting non-relativistic string sigma model on
stringy Newton Cartan geometry was proposed in
\cite{Bergshoeff:2018yvt}
\footnote{For alternative approach to  non-relativistic string in Newton-Cartan background see two 
	recent interesting papers  
\cite{Harmark:2018cdl,Harmark:2017rpg}.}. This is ultraviolet finite theory which
provides a quantization of stringy Newton-Cartan geometry in the
same way as relativistic string theory provides quantum description
of gravity. This model is generalization of the non-relativistic
string theories that were proposed in
\cite{Gomis:2000bd,Danielsson:2000gi}. These theories are invariant
under string Galilean global symmetry and their characteristic (and
crucial) property is existence of two additional world-sheet fields
beyond those parameterizing target space coordinates. The
non-relativistic string sigma model proposed in
\cite{Bergshoeff:2018yvt} couples non-relativistic string to
background fields which are metric, Kalb-Ramond two form field and
dilaton.

Another very interesting result that was found in \cite{Bergshoeff:2018yvt}
is related to the T-duality of the non-relativistic string sigma model on an arbitrary string Newton-Cartan background. Due to the foliation of the stringy Newton-Cartan structure there are two distinct T-duality transformations-transverse and longitudinal. These T-duality transformations were analyzed in the context of Lagrangian formalism
with interesting results derived. It was shown that in the case of T-duality along longitudinal spatial direction we find world-sheet theory that corresponds to relativistic string propagating on a Riemannian manifold with a compact lightlike isometry. In other words non-relativistic string on a stringy Newton-Cartan geometry
with longitudinal spatial isometry can be used for definition of the dynamics of relativistic string on Riemannian manifold with compact light-like isometry  and
possibly could be used for definition of its  discrete light cone quantization (DLCQ). It is well known that DLCQ is fundamental block of Matrix theory description of M-theory \cite{Banks:1996vh,Susskind:1997cw,Seiberg:1997ad}. However DLCQ of string theory is non-trivial. In fact,  previous definition of DLCQ of string theory  was based on subtle limit of compactification on a space-like circle \cite{Seiberg:1997ad,Sen:1997we,Hellerman:1997yu} so that
the result derived in
\cite{Bergshoeff:2018yvt} provides definition of DLCQ of string theory on arbitrary Lorentzian background with lightlike isometry.

This fact is very promising since it opens new intriguing direction in the study of M-theory and its DLCQ description. For that reason we mean that T-duality in the context of non-relativistic string theory on stringy Newton-Cartan background
deserves further study. This is the goal of this paper when we would like to give a canonical description of T-duality of non-relativistic string.  In fact, it is well known that in case of relativistic string, T-duality can be also interpreted as canonical transformation
\cite{Alvarez:1994wj,Alvarez:1994dn}. In this paper we show that the same is true in case of non-relativistic string theory
on stringy Newton-Cartan background. In order to do it we have to use the Hamiltonian form of the model \cite{Bergshoeff:2018yvt}
that it was found in \cite{Kluson:2018grx}. With the help of this Hamiltonian we define T-duality as a canonical transformation. We focus on two physically different T-dualities: T-duality in longitudinal spatial direction  and T-duality in transverse direction which have the same description in the canonical formalism. Performing these T-duality transformations we find T-dual Hamiltonian. Then in order to find explicit form of the transformed background field we derive corresponding Lagrangian and we find agreement with \cite{Bergshoeff:2018yvt} in case of
longitudinal spatial and transverse T-duality transformations. Explicitly, we find that T-duality along longitudinal spatial direction leads to T-dual relativistic string on the background with isometry in light-like direction while in case of T-duality in transverse direction we obtain again non-relativistic string in T-dual background. We mean that this is very nice result that again shows how canonical treatment of T-duality transformation can be powerful.

The structure of this paper is as follows. In the next section
(\ref{second}) we review basic facts about stringy Newton-Cartan
background and string sigma model together with its canonical
formulation. In section (\ref{third}) we introduce T-duality as
canonical transformation, following
\cite{Alvarez:1994wj,Alvarez:1994dn} and perform T-duality
transformation along longitudinal spatial direction and determine
background fields. Then we perform similar analysis in case of
transverse T-duality.

\section{Review of Stringy Newton-Cartan Geometry}\label{second}
Let us define string Newton-Cartan geometry following
\cite{Bergshoeff:2018yvt}. Let $\mM$ is $D+1$ dimensional manifold
and let $\mathcal{T}_p$ is tangent space at point $p$. We decompose
$\mathcal{T}_p$ into longitudinal directions  indexed by $A=0,1$ and
transverse directions with $A'=2,\dots,d-1$. Two dimensional
foliation of $\mM$ is defined by generalized clock function
$\tau_\mu^{ \ A}$ that is also known as longitudinal vielbein field
that satisfies a constraint
\begin{equation}
D_\mu \tau_\nu^{ \ A}-D_\nu\tau_\mu^{ \ A}=0 \ ,
\end{equation}
where $D_\mu$ is covariant derivative with respect to the longitudinal Lorentz
transformations acting on index $A$. Let us also introduce transverse vielbein
field $E_\mu^{ \ A'}$. We further introduce projective inverse $\tau^\mu_{ \ A}$ and
$E^\mu_{ \ A'}$ that are defined as
\begin{equation}
E_\mu^{ \ A'}E^\mu_{\ B'}=\delta^{A'}_{B'} \ , \quad
\tau^\mu_{ \ A}\tau_\mu^{ \ B}=\delta_A^{ B} \  \nonumber \\
\end{equation}
 that obey following relations
\begin{eqnarray}
& & \tau_\mu^{ \ A}\tau^\nu_{ \ A}+E_\mu^{ \ A'}E^\nu_{ \ A'}=\delta^\nu_\mu \ ,
\nonumber \\
& & \tau^\mu_{ \ A}E_\mu^{ \ A'}=0 \ , \quad \tau_\mu^{ \ A}E^\mu_{ \ A'}=0 \ .
\nonumber \\
\end{eqnarray}
From $\tau_\mu^{ \ A}$  we can construct longitudinal metric $\tau_{\mu\nu}=\tau_\mu^{ \ A}\tau_\nu^{ \ B}\eta_{AB}$ and
transverse metric $H^{\mu\nu}=E^\mu_{ \ A'}E^\nu_{ \ B'}\delta^{A' B'}$.

It is clear that in order to  define string moving in stringy Newton-Cartan background we
need tensor $H_{\mu\nu}$. It was shown in  \cite{Bergshoeff:2018yvt} that such
a tensor has the form
\begin{equation}
H_{\mu\nu}=E_\mu^{ \ A'}E_\nu^{ \ B'}\delta_{A'B'}+(\tau_\mu^{ \ A}m_\nu^{ \ B}+
\tau_\nu^{ \ A}m_\mu^{ \ B})\eta_{AB} \ .
\end{equation}
Now we are ready to write  an action for non-relativistic string in this background
\cite{Bergshoeff:2018yvt}. It turns out that this action contains
world-sheet scalars $x^\mu$ that parameterize an embedding string into target space time together with two additional world-sheet fields  that we denote as $\lambda$ and $\blambda$. These fields are needed for the realization of string Galilei symmetry on the world-sheet theory.

Now we will be more explicit. Let $\sigma^\alpha,\sigma^0\equiv \tau \ , \sigma^1\equiv \sigma$ parameterize world-sheet surface $\Sigma$. The sigma model is endowed with two dimensional world-sheet metric $h_{\alpha\beta}$ and we introduce two dimensional vielbein
$e_\alpha^{ \ a} \ , a=0,1$ so that
\begin{equation}
h_{\alpha\beta}=e_\alpha^{ \ a}e_\beta^{ \ b}\eta_{ab} \ .
\end{equation}
Using light cone coordinates for the flat index $a$ on the
world-sheet tangent space we define
\begin{equation}
e_\alpha\equiv e_\alpha^{ \ 0}+e_\alpha^{\ 1} \ , \quad
\bar{e}_\alpha\equiv e_\alpha^{ \ 0}-e_\alpha^{\  1} \ .
\end{equation}
We can also use light-cone coordinates for the flat index $A$ on the
space-time tangent space $\mathcal{T}_p$ and define
\begin{equation}
\tau_\mu\equiv \tau_\mu^{ \ 0}+\tau_\mu^{ \ 1} \ ,
\bar{\tau}_\mu=\tau_\mu^{ \ 0}-\tau_\mu^{\ 1} \ .
\end{equation}
Then we are ready to write sigma model for non-relativistic string on an arbitrary
string Newton-Cartan geometry, and dilaton background in the form
\begin{eqnarray}\label{Saction}
S&=&-\frac{T}{2}\int d^2\sigma (\sqrt{-h}h^{\alpha\beta}
\partial_\alpha x^\mu\partial_\beta x^\nu H_{\mu\nu}+\epsilon^{\alpha\beta}
(\lambda e_\alpha \tau_\mu+\bar{\lambda}\bar{e}_\alpha \bar{\tau}_\mu)
\partial_\beta x^\mu)-\nonumber \\
&-&\frac{T}{2}\int d^2\sigma \epsilon^{\alpha\beta}\partial_\alpha x^\mu
\partial_\beta x^\nu B_{\mu\nu}+\frac{1}{4\pi}\int d^2\sigma \sqrt{-h}R\Phi \ ,
\nonumber \\
\end{eqnarray}
where $h=\det h_{\alpha\beta} \ , h^{\alpha\beta}$ is inverse to
$h_{\beta\alpha}$, $R$ is scalar curvature of $h_{\alpha\beta}$ and $T$ is string tension.
In what follows we restrict to the case of constant dilaton field so that the
last term on the second line is total derivative and will be ignored. In other words, since we restrict ourselves to the classical canonical analysis we cannot determine
transformation rules for dilaton under T-duality transformations.
\subsection{Hamiltonian for String in Stringy Newton-Cartan Geometry}
In our previous work \cite{Kluson:2018grx} we performed Hamiltonian analysis of string in stringy Newton background. We found that the Hamiltonian is sum of two first class constraints $\mH_\tau,\mH_\sigma$ that have the form
\begin{eqnarray}\label{Hconst}
& &\mH_\tau=
\frac{1}{T}
\pi_\mu H^{\mu\nu}\pi_\nu+
Tx'^\mu H_{\mu\nu}x'^\nu-\pi_\mu H^{\mu\nu}
(\lambda^+\tau_\nu+\lambda^-\bar{\tau}_\nu)+
\nonumber \\
& &+Tx'^\mu(\lambda^+\tau_\mu-
\lambda^-\bar{\tau}_\mu)
+\frac{T}{4}(\lambda^+\tau_\mu+\lambda^-\bar{\tau}_\mu)
H^{\mu\nu}(\lambda^+\tau_\nu+\lambda^-\tau_\nu) \ ,
\nonumber \\
& & \mH_\sigma=
x'^\mu p_\mu\ , \quad
\pi_\mu=p_\mu+TB_{\mu\rho}x'^\rho \  , \quad x'^\mu\equiv \partial_\sigma x^\mu \ ,
\nonumber \\
\end{eqnarray}
where the matrix $H^{\mu\nu}$ is inverse to $H_{\mu\nu}$ so that
$H_{\mu\nu}H^{\nu\rho}=\delta_\mu^\rho$ with following explicit form
\begin{equation}
H^{\mu\nu}\equiv
h^{\mu\nu}-\htau^\mu_{ \ A}(\Phi^{-1})^{AB}\htau^\nu_{ \ B} \ ,
\quad
H^{\mu\nu}H_{\nu\rho}=\delta^\mu_\rho  \ ,
\end{equation}
where we introduced  $\htau^\mu_{ \ A}$ defined as
\begin{equation}
\htau^\mu_{ \ A}=\tau^\mu_{ \ A}-h^{\mu\rho}m_\rho^{ \ B}\eta_{BA} \ .
\end{equation}
Further, we also defined $2\times 2$ matrix $(\Phi^{-1})^{AB}$ with following explicit form
\begin{equation}
(\Phi^{-1})^{AB}=\frac{1}{\det\Phi_{AB}}\left(\begin{array}{cc}
\Phi_{11} & -\Phi_{01} \\
-\Phi_{01} & \Phi_{00} \\ \end{array}\right) \ .
\end{equation}
Clearly $(\Phi^{-1})^{AB}$ is the matrix inverse to the matrix valued Newton potential that is defined as
\begin{eqnarray}
\Phi_{AB}=-\tau^\sigma_{ \ A}m_\sigma^{ \ C}\eta_{CB}
-\eta_{AC}m_\rho^{ \ C}\tau^\rho_{ \ B}
+\eta_{AC}m_\rho^{ \ C}h^{\rho\sigma}m_\sigma^{ \ D}\eta_{DB} \ .
\nonumber \\
\end{eqnarray}
Finally, $\lambda^+,\lambda^-$ are two world-sheet scalar fields which
are related to $\lambda,\blambda$, for more details see \cite{Kluson:2018grx}.
Since $\lambda^+,\lambda^-$ are non-dynamical their conjugate momenta are
primary constraints and the requirement of their preservation
implies an existence of two secondary  constraints in the form
\begin{eqnarray}
\mG^\lambda_+=\pi_\mu H^{\mu\nu}\tau_\nu-
Tx'^\mu \tau_\mu-\frac{T}{2}\tau_\mu H^{\mu\nu}(\lambda^+\tau_\nu+\lambda^-\bar{\tau}_\nu)
\approx 0 \ , \nonumber \\
\mG^\lambda_-=\pi_\mu H^{\mu\nu}\bar{\tau}_\nu+
Tx'^\mu\bar{\tau}_\mu-\frac{T}{2}\bar{\tau}_\mu H^{\mu\nu}
(\lambda^+\tau_\nu+\lambda^-\bar{\tau}_\nu)\approx 0 \ .
\nonumber \\
\end{eqnarray}
These constraints are second class constraints together with momenta
conjugate to $\lambda^+,\lambda^-$. Then it was shown  in
\cite{Kluson:2018grx} that these constraints can be solved for
$\lambda^+$ and $\lambda^-$. Plugging these solutions to the
original Hamiltonian constraint we obtain Hamiltonian constraint
corresponding to the non-relativistic string action that was
proposed in \cite{Kluson:2018uss}.

\section{T-duality in Canonical Formalism}\label{third}
In this section we present canonical analysis of T-duality for non-relativistic string.
Due to the non-trivial foliation of target space-time we have to distinguish between
T-duality transformations in the longitudinal and transverse directions. However we will see that these two transformations have the same treatment in the canonical formalism.
Let us start our analysis with  \emph{longitudinal spatial T-duality transformation}.
\subsection{Longitudinal Spatial T-Duality Transformation}
Following \cite{Bergshoeff:2018yvt}  we presume that non-relativistic theory possesses longitudinal spatial Killing vector $k^\mu$ that obeys the relation
\begin{equation}
\tau_\mu^{ \ 0}k^\mu=0 \ , \quad \tau_\mu^{\ 1}k^\mu \neq 0 \ , \quad E_\mu^{ \ A'}k^\mu=0
\ .
\end{equation}
It is convenient to introduce coordinate system $(y,x^i)$ adapted to $k^\mu$ such that $k^\mu \partial_\mu=\partial_y$. It is important to stress that $x^i$ contains
longitudinal coordinate. Then the previous condition implies
\begin{equation}
\tau_y^{ \ 0}=0 \ , \quad  \tau_y^{ \ 1}\neq 0 \ , \quad E_y^{ \ A'}=0 \
, \quad  \tau_y=-\bar{\tau}_y\neq 0 \ .
\end{equation}
Finally, in adapted coordinates all background fields are independent on $y$. As a result the theory is invariant under shift
\begin{equation}
y' \rightarrow y+\epsilon \ , \epsilon=\mathrm{const} \ .
\end{equation}
Our goal is to perform canonical transformation from $y$ to $\ty$ in the same way as in
case of relativistic string  \cite{Alvarez:1994wj,Alvarez:1994dn}. As was shown there
 the
generating function has the form
\begin{equation}
G(y,\ty)=\frac{T}{2}\int d\sigma (\partial_\sigma y\ty-y\partial_\sigma \ty) \ .
\end{equation}
Let us denote momentum conjugate to $\ty$ as $p_{\ty}$. Then from
the definition of the canonical transformations we derive following
relation between $\ty$ and $p_{\ty}$ in the form
\begin{eqnarray}
& &p_{\ty}=-\frac{\delta G}{\delta \ty}=-T\partial_\sigma y \ , \nonumber \\
& &p_y=\frac{\delta G}{\delta y}=-T\partial_\sigma \ty \ . \nonumber \\
\end{eqnarray}
With the help of these relations we obtain dual Hamiltonian when we replace $
\partial_\sigma y$ with $-\frac{1}{T}p_{\ty}$ and $p_y$ with $-T\partial_\sigma y$.
Explicitly, using the constraints given in (\ref{Hconst}) we obtain T-dual constraints
in the form
\begin{eqnarray}\label{HTdual}
& &\mH_\tau^T=
\frac{1}{T}(k_i-B_{iy}p_{\ty})H^{ij}(k_j-B_{jy}p_{\ty})+\frac{2}{T}(-T \ty' +TB_{yk} x'^k)H^{yi}(k_i-B_{i\ty}p_{\ty})+\nonumber \\
& &+\frac{1}{T}(-T\ty'+TB_{yi}x'^i)H^{yy}(-T\ty'+TB_{yj}x'^j)+\frac{1}{T}p_{\ty}H_{yy}p_{\ty}-2p_{\ty}H_{yj}x'^j+Tx'^iH_{ij}x'^j-\nonumber \\
& &-(-T\ty'+TB_{yi}x'^i)H^{yy}(\lambda^+\tau_y+\lambda^-\bar{\tau}_y)-
(-T\ty'+TB_{yi}x'^i)H^{yj}(\lambda^+\tau_j+\lambda^-\bar{\tau}_j)-\nonumber \\
& &-(k_i-B_{iy}p_{\ty})H^{ij}(\lambda^+\tau_j+\lambda^-\bar{\tau}_j)-(k_i-B_{iy}p_{\ty})
H^{iy}(\lambda^+\tau_y+\lambda^-\bar{\tau}_y)
+
\nonumber \\
& &+Tx'^i(\lambda^+\tau_i-
\lambda^-\bar{\tau}_i)-p_{\ty}(\lambda^+\tau_y-\lambda^-\bar{\tau}_y)
+\frac{T}{4}(\lambda^+\tau_\mu+\lambda^-\bar{\tau}_\mu)
H^{\mu\nu}(\lambda^+\tau_\nu+\lambda^-\bar{\tau}_\nu) \ ,
\nonumber \\
& &\mH^T_\sigma=
x'^i p_i+\ty'p_{\ty} \ , \nonumber \\
\end{eqnarray}
where $k_i=p_i+TB_{ij}\partial_\sigma x^j$.  We see that it is very difficult to
find T-dual background fields from this form of the Hamiltonian which is a consequence of the fact that symmetries are usually hidden in the canonical formalism. On the other hand
symmetric structures and forms of the background fields naturally emerge in the Lagrangian formalism so that we should determine Lagrangian density corresponding
to T-dual constraints (\ref{HTdual}). To do this we determine time evolution of $x^i$ and $\ty$ using T-dual Hamiltonian in the form
\begin{equation}
H^T=\int d\sigma (N\mH^T_\tau+N^\sigma \mH^T_\sigma) \ ,
\end{equation}
where $\mH_\tau^T,\mH_\sigma^T$ are given in (\ref{HTdual}). Explicitly we obtain
\begin{eqnarray}
& &\partial_\tau x^i=\pb{x^i,H^T}=
\frac{2N}{T}H^{ij}(k_j-B_{jy}p_{\ty})+\frac{2N}{T}H^{iy}
(-T\ty'+TB_{yj}x'^j)-\nonumber \\
& &-NH^{ij}(\lambda^+\tau_j+\lambda^-\bar{\tau}_j)
-N H^{iy}(\lambda^+\tau_y+\lambda^-\bar{\tau}_y)
+N^\sigma x'^i\ , \nonumber \\
& &\partial_\tau \ty=\pb{\ty,H^T}=
-\frac{2N}{T}B_{iy}H^{ij}(k_j-B_{jy}p_{\ty})-\frac{2N}{T}(-T\ty'+T
B_{yj}x'^j)H^{yi}B_{iy}+\nonumber \\
& &+\frac{2N}{T}H_{yy}p_{\ty}-2NH_{yj}x'^j
+NB_{iy}H^{ij}(\lambda^+\tau_j+\lambda^-\bar{\tau}_j)
+NB_{iy}H^{iy}(\lambda^+\tau_y+\lambda^-\bar{\tau}_y)-\nonumber \\
& &-N(\lambda^+\tau_y-\lambda^-\bar{\tau}_y)+N^\sigma \ty' \ .
\nonumber \\
\end{eqnarray}
From the last equation we can express $p_{\ty}$ as
\begin{equation}
p_{\ty}=\frac{T}{2NH_{yy}}(\tY+B_{jy}X^j+2NH_{yj}x'^j+N(\lambda^+\tau_y-\lambda^-
\bar{\tau}_y)) \ , \nonumber \\
\end{equation}
where we defined
\begin{equation}
X^i=\dot{x}^i-N^\sigma x'^i \ , \quad
\tY=\dot{\ty}-N^\sigma \ty' \ .
\end{equation}
In case of $k_i$ the situation is more involved since we have to find metric inverse to $H^{ij}$. It turns out that it has the form
\begin{equation}
h_{ij}=H_{ij}-\frac{H_{iy}H_{yj}}{H_{yy}} \ .
\end{equation}
For further purposes we write following relation
\begin{equation}
h_{ij}H^{jy}=-\frac{H_{iy}}{H_{yy}}
\end{equation}
that follows from the definition of $h_{ij}$. With the help of the inverse metric
we find corresponding Lagrangian density
\begin{eqnarray}\label{mLTdual}
& &\mL^T=p_{\ty}\partial_\tau\ty+p_i\partial_\tau x^i-H^T=
\nonumber \\
& &=\frac{T}{4N}(\tg_{\sigma\sigma}-2N^\sigma \tg_{\sigma\tau}+(N^\sigma)^2
\tg_{\sigma\sigma})-NT \tg_{\sigma\sigma}-T\tB_{\mu\nu}\partial_\tau \tx^\mu
\partial_\sigma \tx^\nu+\nonumber \\
&&+\frac{T}{2}\lambda^+A+\frac{T}{2}\lambda^-B+\frac{NT}{H_{yy}} \lambda^+\lambda^-\tau_y\tau_y  \ ,
\nonumber \\
\end{eqnarray}
where
\begin{equation}
\tg_{\alpha\beta}=\partial_\alpha \tx^\mu \tH_{\mu\nu}\partial_\beta
\tx^\nu \ , \quad
\tx^\mu=(x^i,\ty) \ ,
\end{equation}
and where we have T-dual components of metric and NSNS two form
\begin{eqnarray}
& &\tH_{yy}=\frac{1}{H_{yy}} \ , \quad \tH_{ij}=H_{ij}-\frac{H_{iy}H_{yj}}{H_{yy}}
+\frac{1}{H_{yy}}B_{iy}B_{jy} \ , \quad
\tH_{yi}=\tH_{iy}=\frac{B_{iy}}{H_{yy}} \ , \nonumber \\
& &\tB_{ij}=B_{ij}+\frac{B_{iy}H_{yj}}{H_{yy}}-\frac{H_{iy}B_{yj}}{H_{yy}} \  , \quad
\tB_{\ty i}=\frac{H_{yi}}{H_{yy}} \ , \quad \tB_{i\ty}=-\frac{H_{yi}}{H_{yy}} \ .
\nonumber \\
\end{eqnarray}
Note that these  transformation components of background fields agree with known Buscher's rules  \cite{Buscher:1987sk,Buscher:1987qj}. The novelty of non-relativistic
Lagrangian is the presence of terms on the last line in
(\ref{mLTdual}) where $A$ and $B$ are equal to
\begin{eqnarray}
A
=\frac{1}{H_{yy}}\left(
\dot{x}^i\bA_i+\dot{\ty}\tau_y-N^\sigma[x'^i\bA_i+\ty' \tau_y]-
2N[x'^i\bA_i+\ty' \tau_y]\right) \ ,
\nonumber \\
B=
\frac{1}{H_{yy}}\left(
\dot{x}^i\bB_i-\dot{\ty}\bar{\tau}_y-N^\sigma[x'^i\bB_i-\ty' \bar{\tau}_y]+
2N[x'^i\bB_i-\ty' \bar{\tau}_y]\right) \ ,
\nonumber \\
\end{eqnarray}
where $\bA_\mu$ and $\bB_\mu$ are equal to
\begin{eqnarray}\label{bAbB}
& &\bA_i=H_{yy}\tau_i+B_{iy}\tau_y-H_{iy}\tau_y \ , \quad \bA_y=\tau_y \ , \nonumber \\
& &\bB_i=H_{yy}\bar{\tau}_i-B_{iy}\bar{\tau}_y-H_{iy}\bar{\tau}_y \ , \quad
\bB_y=-\bar{\tau}_y \ .
\nonumber \\
\end{eqnarray}
As the last step we will solve the equations of motion for $\lambda^+$ and $\lambda^-$ that have the form
\begin{eqnarray}
\frac{1}{2}A+\frac{N}{H_{yy}}\lambda^-\tau_y\tau_y=0 \ , \quad
\frac{1}{2}B+\frac{N}{H_{yy}}\lambda^+\tau_y\tau_y=0
\end{eqnarray}
with corresponding solutions
\begin{equation}
\lambda^-=-\frac{H_{yy}}{2N}\frac{A}{\tau_y\tau_y} \ , \quad
\lambda^+=-\frac{H_{yy}}{2N}\frac{B}{\tau_y\tau_y} \ .
\end{equation}
Inserting back to the Lagrangian density (\ref{mLTdual}) we obtain
\begin{eqnarray}
\mL^T&=&\frac{T}{4N}
(\tg_{\sigma\sigma}-2N^\sigma \tg_{\sigma\tau}+(N^\sigma)^2
\tg_{\sigma\sigma})-NT \tg_{\sigma\sigma}-\nonumber \\
&-&T\tB_{\mu\nu}\partial_\tau \tx^\mu
\partial_\sigma \tx^\nu
-\frac{1}{4N}T\frac{H_{yy}}{\tau_y\tau_y}AB  \ ,
\nonumber \\
\end{eqnarray}
where
\begin{eqnarray}
& &-\frac{H_{yy}}{4N\tau_y\tau_y}AB
=
\nonumber \\
& &-\frac{1}{4N\tau_y\tau_y H_{yy}}
[\frac{1}{2}\partial_\tau\tx^\mu\partial_\tau\tx^\nu(\bA_\mu\bB_\nu+\bA_\nu\bB_\mu)-2N^\sigma \partial_\tau\tx^\mu\partial_\sigma\tx^\nu\frac{1}{2}(\bA_\mu\bB_\nu+\bA_\nu\bB_\mu)+\nonumber \\
& &+2N\partial_\tau\tx^\mu\partial_\sigma\tx^\nu(\bA_\mu \bB_\nu-\bA_\nu\bB_\mu)+\frac{1}{2}
(N^\sigma)^2\partial_\sigma\tx^\mu\partial_\sigma\tx^\nu(\bA_\mu\bB_\nu+\bA_\nu\bB_\mu)
\nonumber \\
& &-4N^2 \partial_\sigma\tx^\mu
\partial_\sigma\tx^\nu \frac{1}{2}(\bA_\mu \bB_\nu+\bA_\nu\bB_\mu)] \ .
\nonumber \\
\end{eqnarray}
Then using explicit form of $\bA_\mu,\bB_\mu$ given in (\ref{bAbB}) we obtain
final form of the  Lagrangian
density
\begin{eqnarray}\label{mLTfinal}
\mL^T&=&\frac{T}{4N}
(g'_{\tau\tau}-2N^\sigma g'_{\sigma\tau}+(N^\sigma)^2
g'_{\sigma\sigma})-NT g'_{\sigma\sigma}
-TB'_{\mu\nu}\partial_\tau \tx^\mu
\partial_\sigma \tx^\nu \ ,
\nonumber \\
\end{eqnarray}
where
\begin{equation}
g'_{\alpha\beta}=H'_{\mu\nu}\partial_\alpha \tx^\mu \partial_\beta \tx^\nu \ ,
\end{equation}
where
\begin{eqnarray}
& &H'_{ij}=H_{ij}+\frac{H_{yy}}{\tau_{yy}}\tau_{ij}-\frac{\tau_{iy}}{\tau_{yy}}
H_{jy}-\frac{\tau_{jy}}{\tau_{yy}}H_{iy}+\frac{1}{\tau_{yy}}
(B_{yi}\tau_j^{ \ A}\tau_y^{ \ B}\epsilon_{AB}+B_{yi}\tau_j^{ \ A} \tau_y^{ \ B}\epsilon_{AB}) \ ,
\nonumber \\
& & H'_{\ty\ty}=0 \ , \quad H'_{\ty i}=-\frac{\tau_i^{ \ A}\tau_y^{ \ B}\epsilon_{AB}}{\tau_{yy}} \ ,
\end{eqnarray}
where we have convention $\epsilon_{01}=-1 \ , \epsilon_{10}=1$.
Finally note that $B'_{\mu\nu}$ is equal to
\begin{eqnarray}
& &B'_{ij}
=B_{ij}-\frac{H_{yy}}{\tau_y\tau_y}\tau_i^{ \ A}
\tau_j^{ \ B}\epsilon_{AB}
+\frac{\tau_{iy}B_{yj}-B_{yi}\tau_{jy}}{\tau_y\tau_y}+
\frac{\tau_i^{ \ A}\tau_y^{ \ B}\epsilon_{AB}H_{jy}-
    \tau_j^{ \ A}\tau_y^{ \ B}\epsilon_{AB}H_{iy}}{\tau_y\tau_y}
\ ,     \nonumber \\
& &B'_{i\ty}
=-\frac{\tau_{iy}}{\tau_y\tau_y} \  \nonumber \\
\end{eqnarray}
These transformation rules agree with the result derived in
\cite{Bergshoeff:2018yvt}. We also see that the T-dual string corresponds to
the relativistic string with light-like isometry due to the absence of the
metric component $H'_{\ty\ty}$. To see more explicitly that we have relativistic string
let us solve the equations of motion for $N$ and $N^\sigma$:
\begin{eqnarray}
-g'_{\sigma\tau}+N^\sigma g'_{\sigma\sigma}=0 \ ,
\quad
-\frac{1}{4N^2}(g'_{\sigma\sigma}-2N^\sigma g'_{\sigma\tau}+(N^\sigma)^2
g'_{\sigma\sigma})-g'_{\sigma\sigma}=0
\  .
 \end{eqnarray}
These equations can be solved for $N^\sigma$ and $N$ as
\begin{equation}
N^\sigma=\frac{g'_{\sigma\tau}}{g'_{\sigma\sigma}} \ ,
\quad N=\frac{1}{2}\frac{\sqrt{-\det g'_{\alpha\beta}}}{g'_{\sigma\sigma}}
\end{equation}
Inserting back to (\ref{mLTfinal})
we obtain
\begin{equation}
\mL^T=-T\sqrt{-\det g'_{\alpha\beta}}-TB'_{\mu\nu}\partial_\tau \tx^\mu
\partial_\sigma \tx^\nu \
\end{equation}
that is relativistic string action in Nambu-Goto form.
\subsection{Transverse T-Duality}
Finally we consider transverse T-duality transformations when we presume that the background has transverse symmetry with following  transverse
spatial Killing vector $p^\mu$ that obeys
\begin{equation}
\tau_\mu^{ \ A}p^\mu=0 \ , \quad E_\mu^{ \ A'}p^\mu\neq 0 \ .
\end{equation}
We define coordinate system $x^\mu(x^i,z)$ adapted to $p^\mu$ such that $p^\mu\partial_\mu=\partial_z$. Then isometry defined above implies
\begin{equation}
\tau_z^{ \ A}=0 \ , \quad E_z^{ \ A'}\neq 0
\end{equation}
and consequently
\begin{equation}
\tau_z=\bar{\tau}_z=0 \ , \quad  H_{zz}\neq 0 \ .
\end{equation}
It is also clear that all background fields do not depend on $z$.
As in previous section we perform replacement
\begin{equation}
p_{\tz}=-T
\partial_\sigma z \ , \quad p_z=-T\partial_\sigma \tz
\end{equation}
so that the constraints (\ref{Hconst}) have the form
\begin{eqnarray}
& &\mH_\tau^T=\frac{1}{T}(k_i-B_{iz}p_{\tz})H^{ij}(k_j-B_{jz}p_{\tz})+
\frac{2}{T}(-T\tz'+TB_{z k}x'^k)H^{z i}(k_i-B_{iz}p_{\tz})
+\nonumber \\
& &+\frac{1}{T}(-T\tz'+T B_{z i}x^i)H^{zz}(-T \tz'+TB_{zj}x'^j)
+\frac{1}{T}p_{\tz}H_{zz}p_{\tz}-2p_{\tz}H_{zi}x'^i+Tx'^i H_{ij}x'^j-
\nonumber \\
& &-(k_i-B_{i\tz}p_{\tz})H^{ij}(\lambda^+\tau_j+\lambda^-\bar{\tau}_j)-
(-T\tz'+TB_{zi}x'^i)H^{zj}(\lambda^+\tau_j+\lambda^-\bar{\tau}_j)
+\nonumber \\
& &+Tx'^i(\lambda^+\tau_i-\lambda^-\bar{\tau}_i)+\frac{T}{4}(\lambda^+\tau_i+\lambda^-\bar{\tau}_i)H^{ij}
(\lambda^+\tau_j+\lambda^-\bar{\tau}_j) \ ,
\nonumber \\
& &\mH_\sigma^T=p_ix'^i+p_{\tz}\tz'\ . \nonumber \\
\end{eqnarray}
Then following the same procedure as  in previous section we find the Lagrangian in the form
%
\begin{eqnarray}\label{mLTT}
\mL^T&=&\frac{T}{4N}(\tg_{\tau\tau}-2N^\sigma \tg_{\tau\sigma}+(N^\sigma)^2
\tg_{\sigma\sigma}^2) -NT\tg_{\sigma\sigma}
-T \partial_\tau\tx^\mu\tB_{\mu\nu}\partial_\sigma \tx^\nu+\nonumber \\
&+&\frac{T}{2}\lambda^+(\partial_\tau\tx^\mu\tau_\mu-N^\sigma \partial_\sigma\tx^\mu \tau_\mu-2N
\partial_\sigma \tx^\mu \tau_\mu)+\nonumber \\
&+&\frac{T}{2}\lambda^-(\partial_\tau\tx^\mu\bar{\tau}_\mu-N^\sigma \partial_\sigma \tx^\mu \bar{\tau}_\mu+2N
\partial_\sigma\tx^\mu \bar{\tau}_\mu) \ , \nonumber \\
\end{eqnarray}
where again
\begin{eqnarray}
\tg_{\alpha\beta}=\tH_{\mu\nu}\partial_\alpha \tx^\mu \partial_\beta \tx^\nu \ ,
\end{eqnarray}
where $\tx^\mu=(x^i,\tz)$
and where
\begin{eqnarray}\label{transverseT}
\tH_{ij}=H_{ij}-\frac{H_{iz}H_{zj}}{H_{zz}}
-\frac{B_{iz}B_{zj}}{H_{zz}} \ , \quad
\tH_{i\tz}=\frac{B_{iz}}{H_{zz}} \ , \quad
\tH_{\tz\tz}=\frac{1}{H_{zz}} \ ,
\nonumber \\
\tB_{ij}=B_{ij}+\frac{B_{zi}H_{zj}-H_{iz}B_{zj}}{H_{zz}} \ , \quad
\tB_{i\tz}=\frac{H_{iz}}{H_{zz}} \ , \quad \tB_{z i}=-\frac{H_{zi}}{H_{zz}}  \ .
\nonumber \\
\end{eqnarray}
In other words, the background fields $\tau_\mu$ and $\bar{\tau}_\mu$ do not change. The form of the Lagrangian density given above suggests that the transverse T-duality maps non-relativistic string to the non-relativistic string in T-dual background where the background fields are given
by standard Buscher's rules (\ref{transverseT}) . In order to see that (\ref{mLTT}) describes non-relativist string let us solve the equations of motion for $\lambda^+$ and $\lambda^-$
\begin{eqnarray}\label{eqtr}
\tau_\tau-N^\sigma \tau_\sigma-2N
 \tau_\sigma=0 \ , \quad \bar{\tau}_\tau-N^\sigma  \bar{\tau}_\sigma+2N
 \bar{\tau}_\sigma=0  \ ,
\end{eqnarray}
where $\tau_\alpha\equiv \tau_\mu\partial_\alpha \tx^\mu \ , \bar{\tau}_\alpha=
\bar{\tau}_\mu\partial_\alpha \tx^\mu$. Taking the sum and difference of these two equations we obtain two equations
\begin{eqnarray}
N^\sigma\tau_\sigma^{ \ 0}+2N\tau_\sigma^{  \ 1}=\tau_\tau^{ \ 0} \ , \quad
N^\sigma \tau_\sigma^{ \ 1}+2N\tau_\sigma^{ \ 0}=\tau_\tau^{ \ 1} \
\end{eqnarray}
that can be solved for $N$ and $N^\sigma$ as
\begin{equation}\label{Nsol}
N^\sigma=\frac{\tau_{\tau\sigma}}{\tau_{\sigma\sigma}} \ , \quad N=\frac{1}{2}\frac{\sqrt{-\det \tau_{\alpha\beta}}}{\tau_{\sigma\sigma}} \ .
\end{equation}
Inserting (\ref{Nsol}) into (\ref{mLTT}) and using (\ref{eqtr}) we obtain Lagrangian
density in the form
\begin{equation}
\mL^T=-\frac{T}{2}\sqrt{-\det\tau_{\alpha\beta}}\tau^{\alpha\beta}\tg_{\alpha\beta}-
B'_{\mu\nu}\partial_\tau \tx^\mu\partial_\beta \tx^\nu \ ,
\end{equation}
where $\tau^{\alpha\beta}$ is inverse to $\tau_{\alpha\beta}$. This form of Lagrangian density corresponds to the non-relativistic string action
\cite{Andringa:2012uz} in the background fields given in
(\ref{transverseT}) which is again in agreement with
\cite{Bergshoeff:2018yvt}.

\acknowledgments{This  work  was
    supported by the Grant Agency of the Czech Republic under the grant
    P201/12/G028. }


\end{document}